\documentclass[a4paper]{spie}  

\usepackage{amsmath,amsfonts,amssymb}
\usepackage{graphicx}
\usepackage[colorlinks=true, allcolors=blue]{hyperref}

\title{Compton Polarimeter Prototype for the CUbesat Solar Polarimeter (CUSP) mission}

\author[a]{Nicolas~De~Angelis}
\author[a,b]{Andrea~Alimenti}
\author[c,d]{Riccardo~Campana}
\author[e]{Mauro~Centrone}
\author[c]{Giovanni~De~Cesare}
\author[a]{Ettore~Del~Monte}
\author[a]{Sergio~Di~Cosimo}
\author[a]{Enrico~Costa}
\author[a]{Sergio~Fabiani}
\author[a]{Abhay~Kumar}
\author[a]{Pasqualino~Loffredo}
\author[a]{Giovanni~Lombardi}
\author[f]{Gabriele~Minervini}
\author[a]{Fabio~Muleri}
\author[g]{Paolo~Romano}
\author[a]{Alda~Rubini}
\author[a,b]{Enrico~Silva}
\author[a]{Paolo~Soffitta}
\author[h]{Giovanni~Cucinella}
\author[h]{Vito~Di~Bari}
\author[h]{Simone~Di~Filippo}
\author[h]{Andrea~Negri}
\author[h]{Massimo~Perelli}
\author[i]{Davide~Albanesi}
\author[i]{Valerio~Campamaggiore}
\author[i]{Giulia~de~Iulis}
\author[i]{Andrea~Del~Re}
\author[i]{Paolo~Leonetti}
\author[i]{Alessandro~Zambardi}
\author[j]{Ilaria~Baffo}
\author[j]{Pierluigi~Fanelli}
\author[k]{Costantino~Zazza}
\author[l]{Andrea~Curatolo}
\author[m]{Nicolas~Gagliardi}
\author[l,m]{Dario~Modenini}
\author[l]{Daniele~Pecorella}
\author[m]{Alice~Ponti}
\author[l,m]{Paolo~Tortora}
\author[n]{Valerio~Vagelli}
\author[n]{Daniele~Brienza}
\author[n]{Immacolata~Donnarumma}
\author[n]{Matteo~Mergè}
\author[n]{Emanuele~Zaccagnino}
\author[n]{Alessandro~Turchi}

\affil[a]{INAF-IAPS\\ via del Fosso del Cavaliere 100, 00133 Rome, Italy}
\affil[b]{Department of Industrial, Electronic and Mechanical Engineering, "Roma Tre" University, via V. Volterra 62, 00146 Rome, Italy}
\affil[c]{INAF-OAS Bologna\\ via Gobetti 93/3, 40129 Bologna, Italy}
\affil[d]{INFN Sezione di Bologna, viale Berti Pichat 6/2, 40127 Bologna, Italy}
\affil[e]{INAF-OAR\\ via Frascati 33, 00040, Monte Porzio Catone, Italy}
\affil[f]{INAF-Headquarters\\ viale del Parco Mellini 84, 00136 Rome, Italy}
\affil[g]{INAF-OACT\\ Via S. Sofia 78, 95123, Catania, Italy}
\affil[h]{IMT s.r.l.\\ via Carlo Bartolomeo Piazza 30, 00161 Rome, Italy}
\affil[i]{DEDA Connect s.r.l.\\ via Vincenzo Lamaro 51, 00173 Rome, Italy}
\affil[j]{DEIM, University of Tuscia, Largo dell’Università, 01100 Viterbo, Italy}
\affil[k]{DIBAF, University of Tuscia, Largo dell’Università, 01100 Viterbo, Italy}
\affil[l]{Department of Industrial Engineering - Alma Mater Studiorum Università di Bologna, Via Montaspro 97, 47121 Forlì, Italy}
\affil[m]{Interdepartmental Centre for Industrial Aerospace Research - Alma Mater Studiorum Università di Bologna, Via Carnaccini 12, 47121 Forlì, Italy}
\affil[n]{ASI, via del Politecnico snc\\ 00133 Rome, Italy}

\authorinfo{Further author information: (Send correspondence to N.D.A.)\\N.D.A.: E-mail: \href{mailto:nicolas.deangelis@inaf.it}{nicolas.deangelis@inaf.it}}

\begin{document} 
\maketitle

\begin{abstract}

The space-based CUbesat Solar Polarimeter (CUSP) mission aims to measure the linear polarization of solar flares in the hard X-ray band (25-100~keV) by means of a dual-phase Compton polarimeter. CUSP will allow to study the magnetic reconnection and particle acceleration in the flaring magnetic structures of our star with its unprecedented sensitivity to solar flare polarization. CUSP is a project under development as part of the Alcor Program of the Italian Space Agency aimed at developing new CubeSat missions.\\
    
In the frame of CUSP's Phase B, which started in December 2024, a flight-representative prototype of the Compton polarimeter has been developed and characterized with hard X-ray sources in the laboratory. This prototype consists of a 4$\times$4 central matrix of plastic scintillator bars surrounded by 4 strips of 8 elongated GAGG scintillators, respectively coupled to a multi-anode photomultiplier tube and arrays of avalanche photodiodes. These sensors are read out by custom front-end electronics based on MAROC-3A and SKIROC-2A ASICs with a Xilinx Artix 7 FPGA. The plastic scintillators act as scatterers, while the GAGG bars fully absorb the scattered photons. Coincident plastic-GAGG events allow for reconstructing the Compton scattering direction, whose distribution allows for inferring the polarization parameters of the source.  We report here the measured performance of the polarimeter prototype using well-known radioactive isotopes and X-ray tubes, allowing us to assess the performance of our polarimeter prototype over the full 25-100~keV energy range.

\end{abstract}

\keywords{CUSP, X-ray Polarimetry, Solar Flares, Space Weather, Solar Physics, CubeSat, Instrument Calibration, Organic Scintillators, GAGG Scintillators, Photomultiplier Tubes, Avalanche Photodiodes}

\section{INTRODUCTION: SOLAR FLARE HARD X-RAY POLARIMETRY}
\label{sec:intro}

Solar activity, and in particular solar flares (SFs), is a major driver of space weather, with the potential to disrupt technological activities both in space and on the ground. Flares are often associated with Coronal Mass Ejections (CMEs) and Solar Energetic Particle (SEP) events, but can also occur in isolation, directly accelerating particles towards the Earth. The most powerful eruptive events are typically associated with the most powerful flares: hard X-rays (HXRs) are linked to the particle acceleration that drives CME initiation, so that HXR polarimetry has the potential to improve our knowledge of the initial conditions of the most energetic CME eruptions, while soft X-rays (SXRs) are related to the subsequent CME velocity evolution.\cite{chen2020,temmer2016}

Flares are thought to originate from magnetic reconnection in loop structures within the solar corona, converting stored magnetic energy into particle acceleration, plasma heating, and radiation across the electromagnetic spectrum. The hard X-ray energy spectrum of a flare is generally described by two components: a thermal Bremsstrahlung component, due to plasma heating and expected to be only weakly polarized,\cite{emslie1980} which dominates below $\sim$10~keV together with atomic emission lines, and a non-thermal Bremsstrahlung component produced at the loop top and footpoints by field-aligned accelerated particles, expected to be significantly polarized above $\sim$10-20~keV.\cite{zharkova2010} The relative weight and time evolution of these two components through the different phases of a flare have been characterized observationally in several statistical and case studies.\cite{sainthilaire2008,grigis2004,nagasawa2022} Non-thermal X-rays therefore carry direct information on the distribution, anisotropy, and beaming of the accelerated particle population.\cite{aschwanden2005}

Spectroscopic and imaging observations alone are often insufficient to break the degeneracies between competing models of particle beaming and magnetic field geometry. Linear X-ray polarimetry provides a powerful, complementary diagnostic: the degree (PD) and angle (PA) of linear polarization depend sensitively on the magnetic field geometry, the directionality of the electron beams, and the viewing angle, even in the absence of imaging of the flaring region.\cite{jeffrey2020} To date, however, X-ray polarization measurements of solar flares have remained scarce and statistically limited, yielding mostly upper limits or marginal detections, from the earliest attempts with Intercosmos-4 and OSO-8\cite{tindo1970,tindo1972,intercosmos1972,tramiel1984} to the more recent RHESSI results.\cite{boggs2006,suarezgarcia2006} The CUSP mission aims to address this observational gap by delivering dedicated, high-sensitivity measurements of solar flare hard X-ray polarization in the 25-100~keV range.

Photons interact with matter through three main processes -- the photoelectric effect, Compton scattering, and pair production -- each dominant in a different energy range. Above 20~keV, in the energy band relevant for the non-thermal emission of solar flares, Compton scattering is the dominant interaction process, and it is intrinsically sensitive to the polarization of the incoming photon: the differential cross-section is modulated as $d\sigma/d\Omega \propto 1+\mu\cos(2\phi)$, where $\phi$ is the azimuthal scattering angle with respect to the polarization vector.\cite{kleinnishina1929} Measuring the azimuthal scattering direction of many photons from a source therefore allows the reconstruction of its polarization degree (related to the amplitude of the modulation) and polarization angle (related to its phase). A segmented array of scintillators, as depicted in Figure \ref{fig:compton_cusp}, in which the incoming photon Compton-scatters in one segment (scatterer) and is subsequently absorbed in a different segment (absorber), can be used to measure this azimuthal scattering angle distribution -- the so-called modulation curve -- and thus to infer the polarization parameters of the source.

\newpage
\section{THE CUSP MISSION}
\label{sec:cusp}

The CUbesat Solar Polarimeter (CUSP) is developed in the framework of the Alcor program, a national program of the Italian Space Agency (ASI) dedicated to the development of CubeSat technologies and missions. The collaboration is organized as follows: INAF-IAPS (Rome) acts as prime contractor, holds the PI-ship of the mission, and is responsible for the payload; INAF-OAS (Bologna) is responsible for the detector simulations, and INAF-OAR (Rome) for laboratory software support; IMT s.r.l. provides the satellite platform; DEDA Connect s.r.l. develops the payload electronics; the University of Bologna (CIRI AERO) is responsible for mission analysis; the University of Tuscia hosts and operates the ground segment; and ASI oversees project control.

\begin{figure}[htb]
\centering
\includegraphics[height=0.34\linewidth]{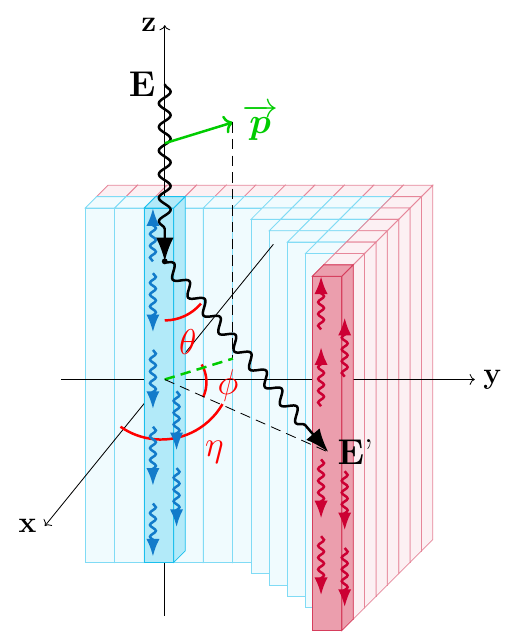}\includegraphics[height=0.34\linewidth]{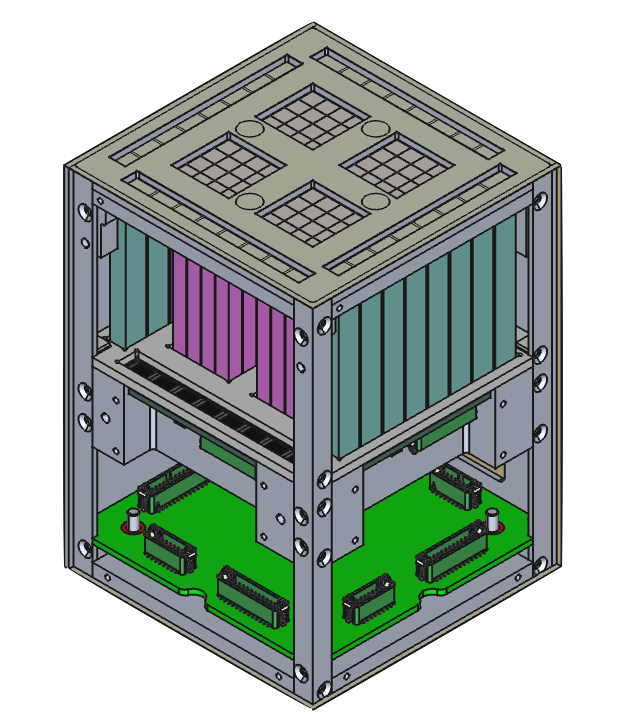}\includegraphics[height=0.34\linewidth]{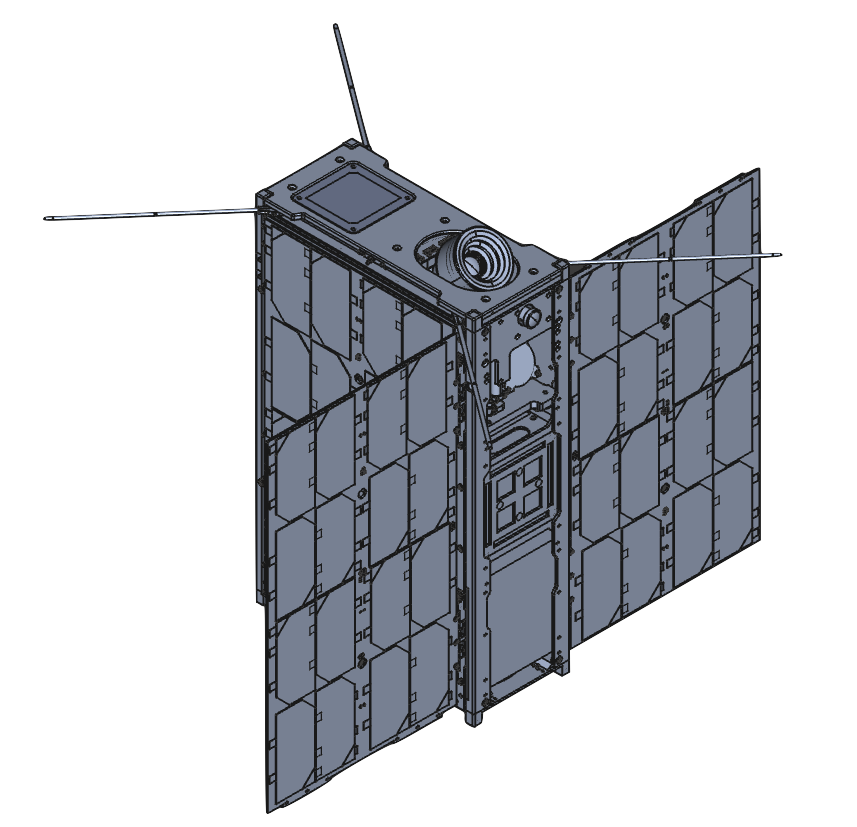}
\caption{\textbf{Left:} Working principle of a Compton scattering polarimeter (Adapted from \cite{DeAngelis:2021v1}, with permission). \textbf{Center:} CAD design of the CUSP payload. \textbf{Right:} 6U-XL platform hosting the Compton polarimeter.}
\label{fig:compton_cusp}
\end{figure}

CUSP consists of a 6U-XL CubeSat platform, already developed by IMT s.r.l. for other projects, hosting a payload of about 2.5~U dedicated to hard X-ray polarimetry of solar flares in the 25-100~keV band. Figure \ref{fig:compton_cusp} shows the CAD design of the payload as well as that of the satellite structure. The satellite will be placed in a Sun-synchronous orbit (SSO) at an altitude of about 500-600~km, from which the Sun can be monitored for approximately 45\% of the time over the nominal 3-year mission lifetime. During Sun-pointed observations, the spacecraft is spun at 1~RPM around the polarimeter symmetry axis: this rotation is designed to average out and reduce the systematic effect known as spurious polarization, arising from residual non-uniformities in the detector response.

The payload is a dual-phase Compton scattering polarimeter (see Figure~\ref{fig:compton_cusp}) that exploits coincident measurements between a central scattering array of $8\times8$ plastic scintillator bars, read out by four 16-channel Multi-Anode PhotoMultiplier Tubes (MAPMTs) from Hamamatsu, and four surrounding strips of 8 elongated GAGG scintillators, acting as absorbers, read out by Avalanche PhotoDiodes (APDs) from Hamamatsu. The MAPMT and APD channels are respectively read out by the MAROC-3A and SKIROC-2A front-end ASICs from Weeroc. As described in Section~\ref{sec:intro}, coincident scatterer-absorber events allow the reconstruction of the Compton scattering direction of each detected photon; the resulting azimuthal distribution over many events (the modulation curve) directly encodes the polarization degree and angle of the source.

The CUSP project completed its 19-month Phase B in July 2026. During this phase a polarimeter prototype was designed, built, and characterized in the laboratory, as reported in Section~\ref{sec:polarimeter}. The mission is now entering a proposed 30-month combined Phase C/D/E1, with a current launch window foreseen for late 2029/early 2030 (see Section~\ref{sec:conclusions}).

\newpage
\section{POLARIMETER PROTOTYPE DESIGN AND TESTING}
\label{sec:polarimeter}

\subsection{Prototype Assembly}

CUSP employs two independent acquisition chains for the scatterer and absorber elements of the polarimeter, respectively based on plastic scintillator bars coupled to MAPMTs and read out by the MAROC-3A ASIC, and on GAGG scintillators coupled to APDs and read out by the SKIROC-2A ASIC. The characterization of these two chains at the single-channel level, performed with development boards and radioactive sources over the 25-100~keV band, is reported in full detail in \cite{deangelis2025spie, deangelisparticles}.

\begin{figure}[htb]
\centering
\includegraphics[height=0.36\linewidth]{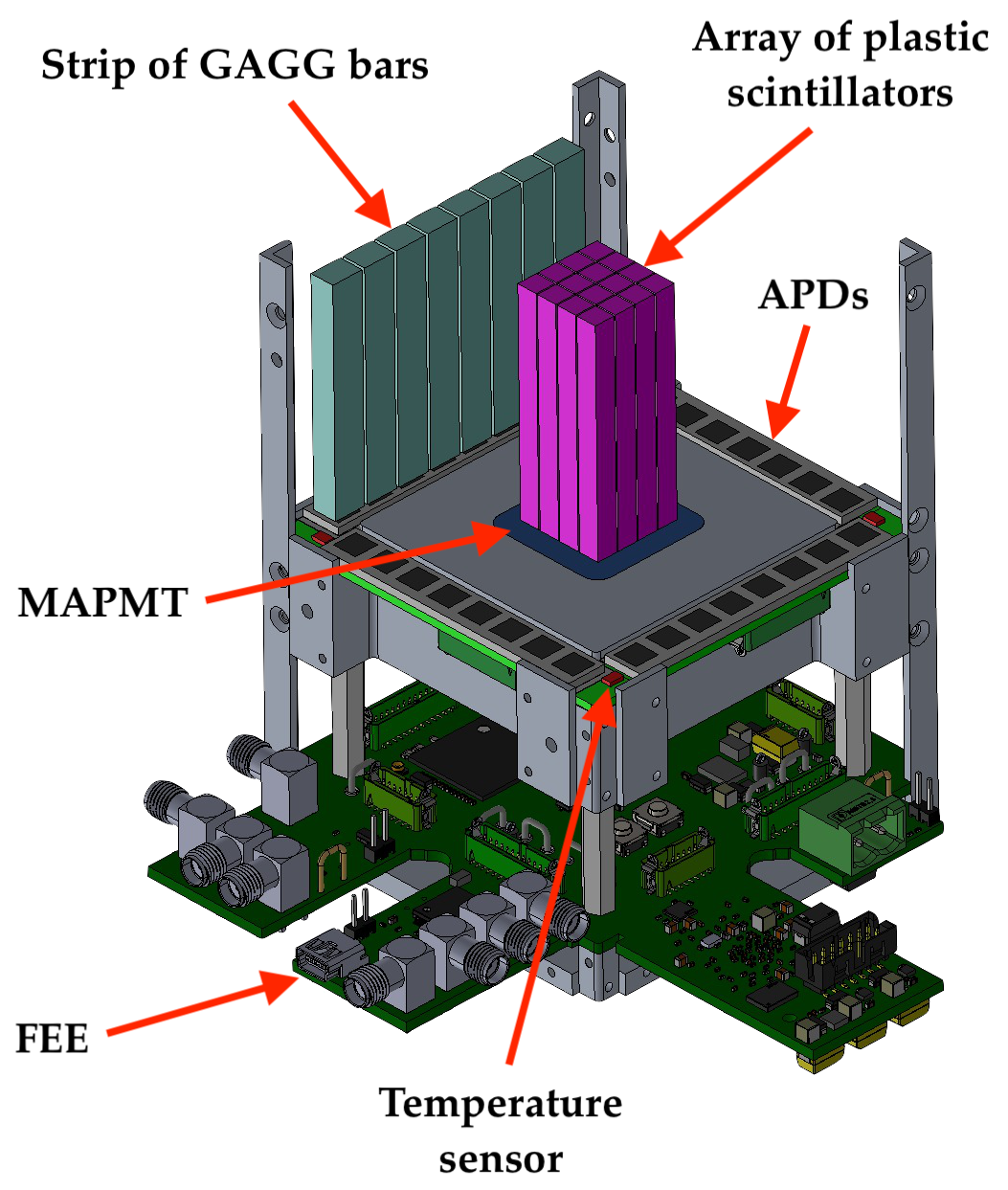}\hspace*{0.2cm}\includegraphics[height=0.36\linewidth]{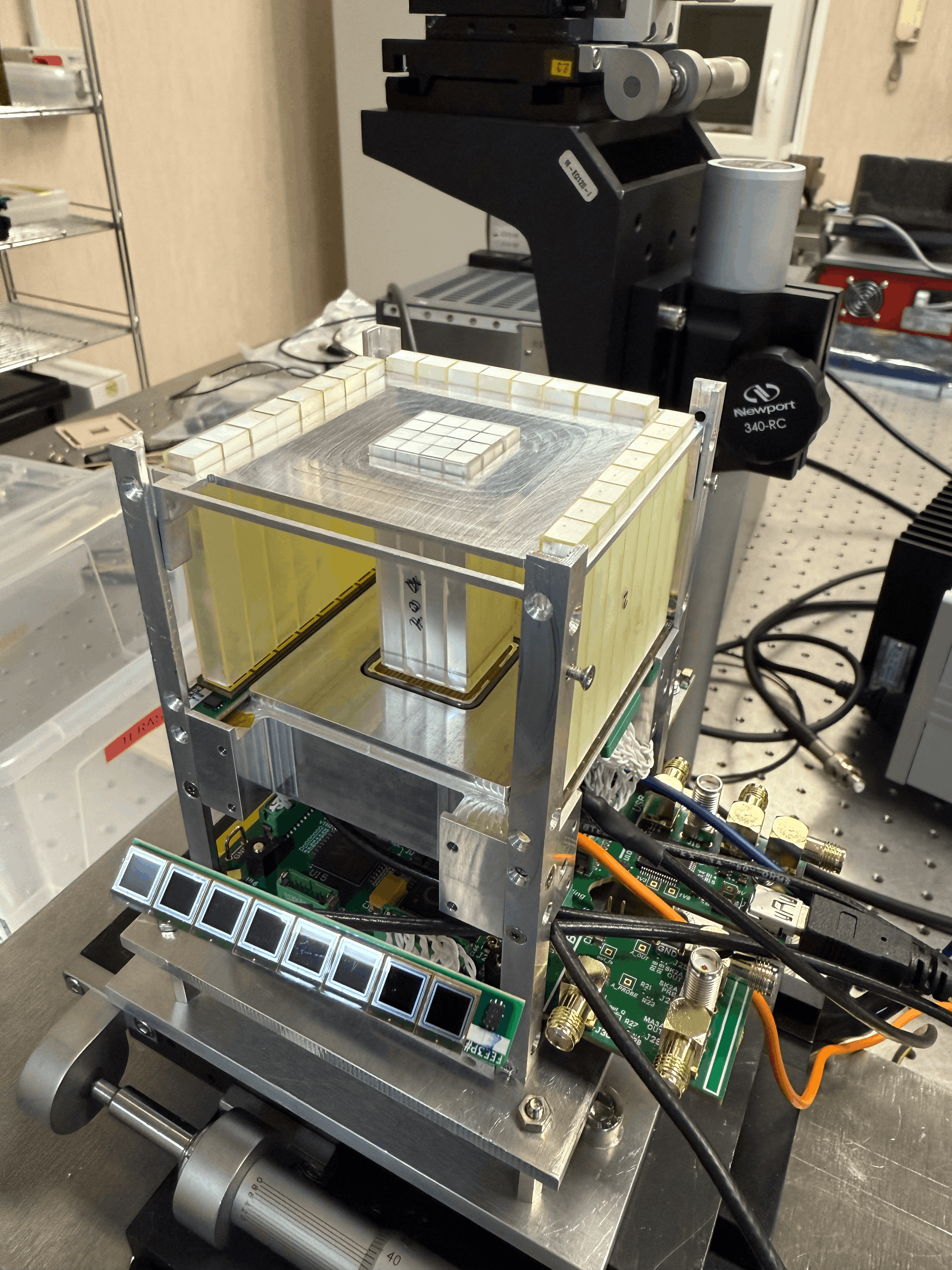}\hspace*{0.2cm}\includegraphics[height=0.36\linewidth]{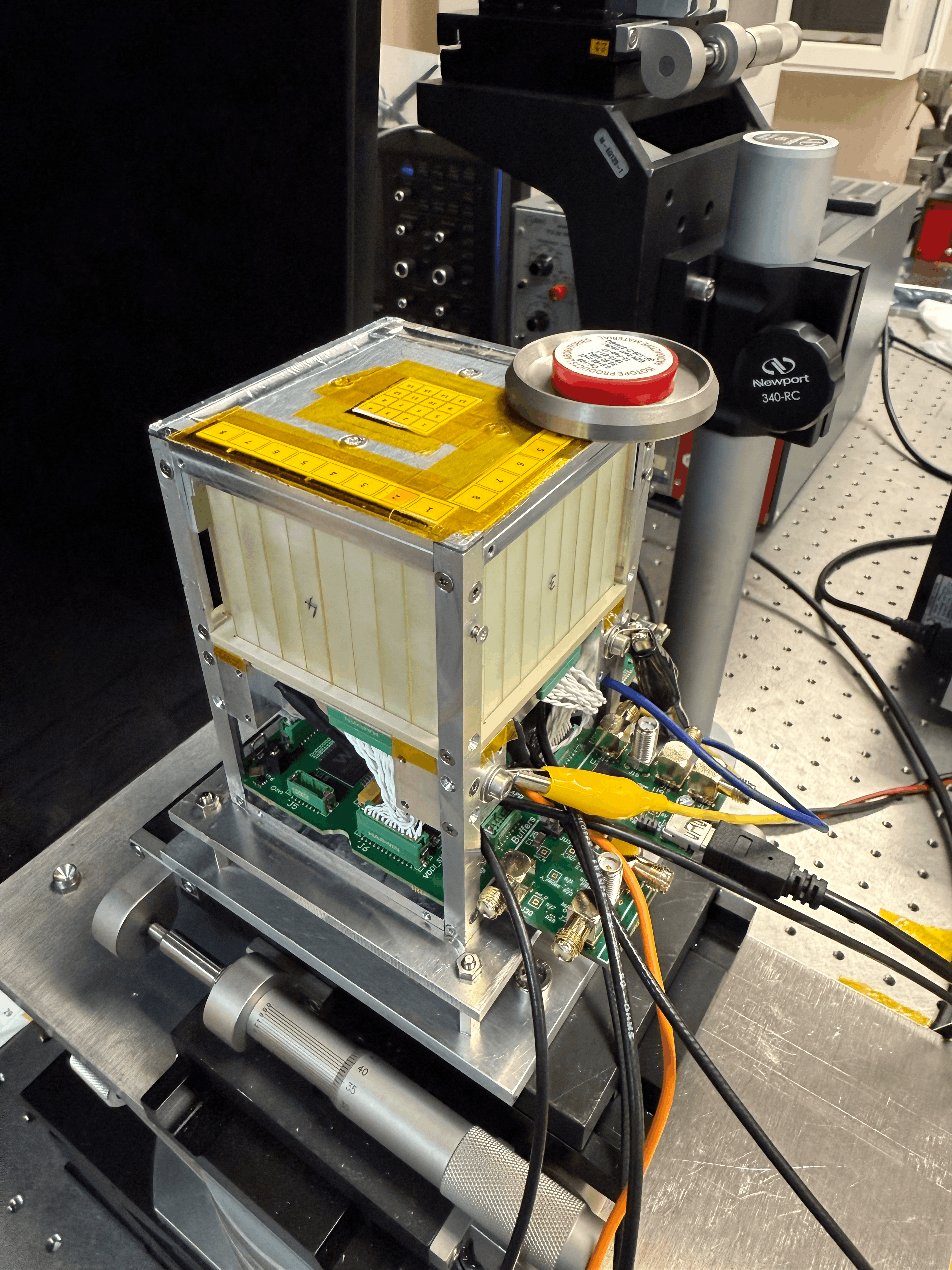}
\caption{\textbf{Left:} Partial CAD drawing of the CUSP prototype (taken from \cite{deangelisparticles}, with permission). \textbf{Center:} Polarimeter prototype being integrated. Three arrays of GAGG scintillators and the central matrix of plastic scintillators are already coupled to the sensors, while the fourth bare strip of APDs can be seen in the foreground. \textbf{Right:} Integrated prototype with a collimated $^{109}$Cd source placed on top of a GAGG scintillator.}
\label{fig:prototype}
\end{figure}

Building on those results, a first polarimeter prototype was assembled and tested by the end of Phase B, as shown in Figure~\ref{fig:prototype}. The prototype consists of a central scattering block, formed by a $4\times4$ matrix of plastic (EJ-204) scintillator bars coupled to an MAPMT, surrounded by GAGG absorber bars coupled to APD arrays, and read out by the flight-representative front-end electronics based on the MAROC-3A and SKIROC-2A ASICs and a Xilinx Artix~7 FPGA.

\subsection{Scatterers Energy Calibration}
\label{sec:scatterer_subsec}

\begin{figure}[htb]
\centering
\includegraphics[height=.38\textwidth]{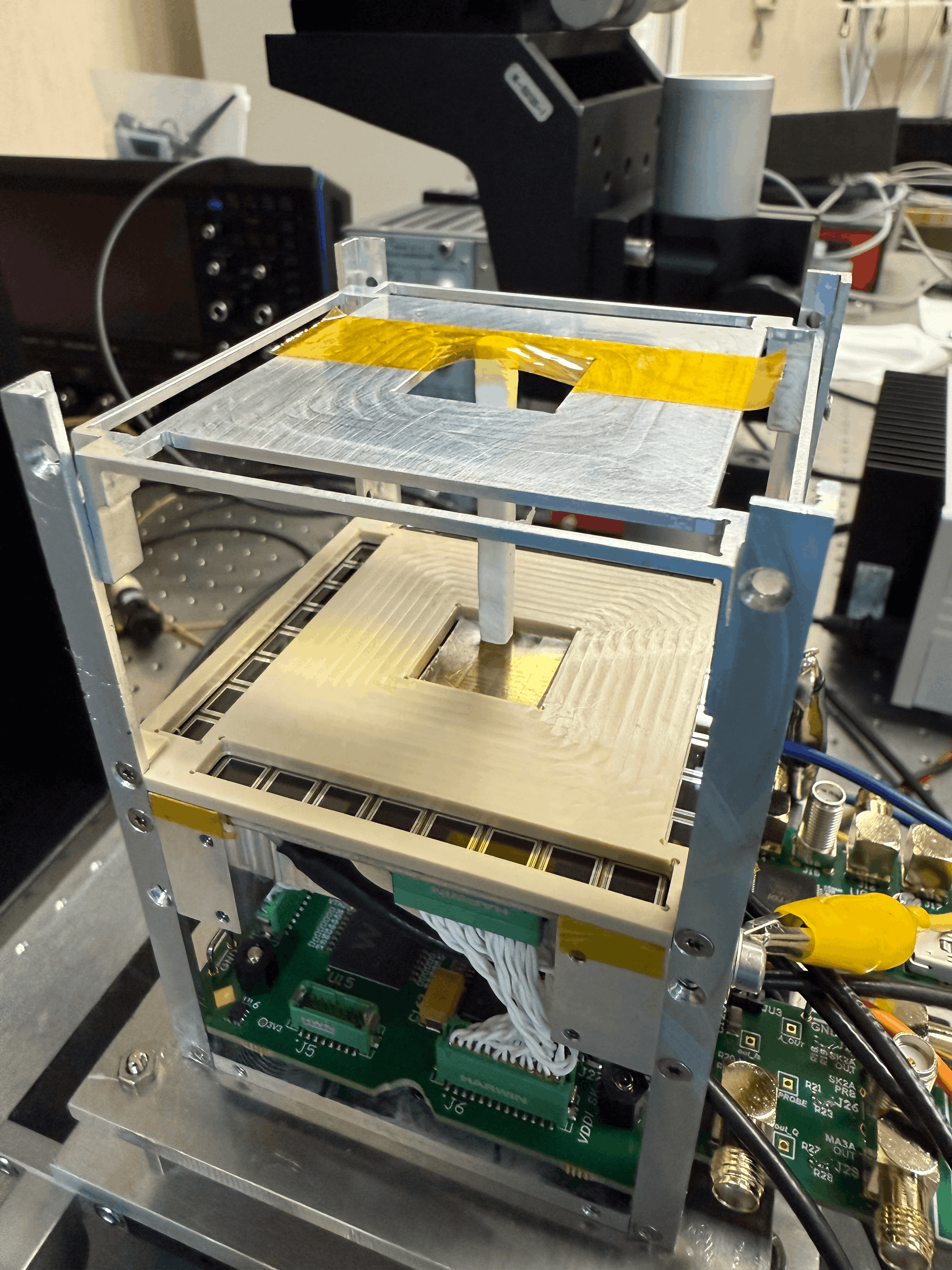}\hspace*{0.2cm}\includegraphics[height=.38\textwidth]{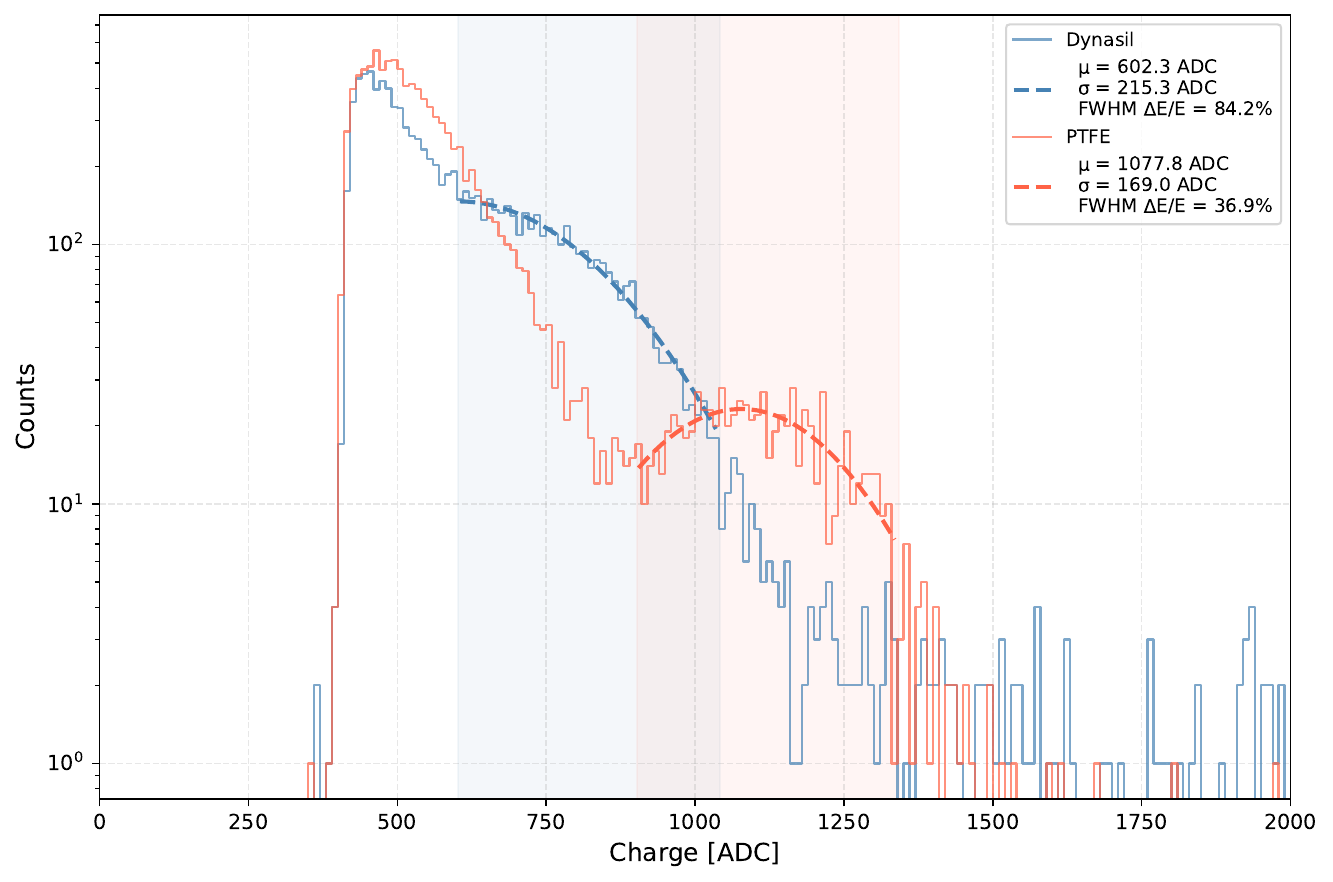}
\caption{\textbf{Left:} Single plastic scintillator bar wrapped with PTFE tape and coupled to a channel of the prototype. \textbf{Right:} Comparison of the $^{241}$Am spectra measured with a single PTFE-wrapped bar and with the plastics array assembled for the prototype.}
\label{fig:plastics_lightyield}
\end{figure}

An energy calibration of the individual plastic scatterer channels was attempted using $^{241}$Am, $^{109}$Cd, and $^{57}$Co radioactive sources. However, none of the characteristic lines from these sources could be resolved above the noise level when using the $4\times4$ plastic scintillator matrix assembled for the prototype, pointing to a strongly reduced light yield of the coupled scintillator-MAPMT chain with respect to the single-channel results reported in \cite{deangelis2025spie, deangelisparticles}. This was traced to the ESR reflective foils having been glued directly onto the scintillator bars, which prevents the total internal reflection normally responsible for guiding the scintillation light to the sensor. The effect was confirmed with a dedicated comparison test on single PTFE-wrapped plastic bars, illuminated with a collimated $^{241}$Am source and read out by the same MAPMT channel (as shown in Figure~\ref{fig:plastics_lightyield}): the 59.5~keV photopeak of $^{241}$Am is clearly resolved above the noise floor with a bare EJ-204 bar wrapped in-house with PTFE tape, but remains buried in the noise for the matrix assembly.

These results indicate that the wrapping procedure of the plastic scatterers will need to be optimized in order to reach the energy resolution required for the mission, and more generally highlight how crucial an accurate understanding and control of light collection is for this kind of segmented scintillator-based instrument. A thorough optical characterization and simulation of the scintillator assemblies, of the kind presented in \cite{deangelis2025jinst}, is therefore essential to correctly inform the optical design of the CUSP scatterers ahead of the larger-scale prototype.

\subsection{Absorbers Energy Calibration}

\begin{figure}[htb]
\centering
\includegraphics[height=.35\textwidth]{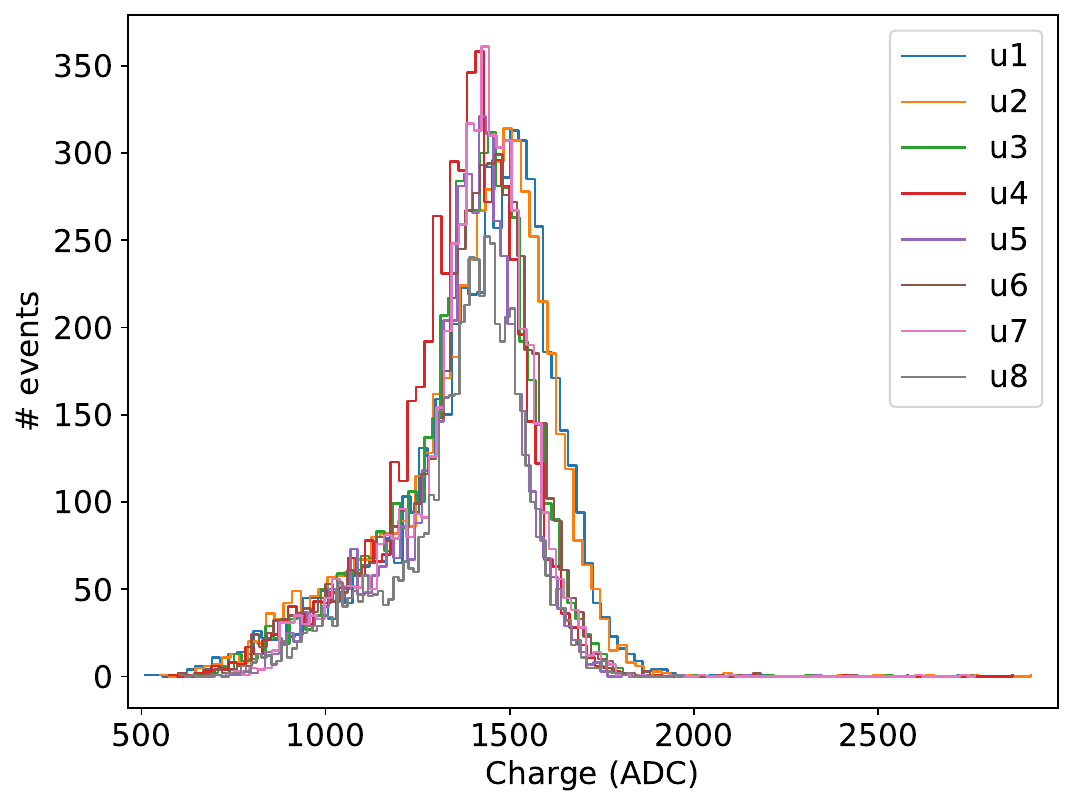}\hspace*{0.2cm}\includegraphics[height=.35\textwidth]{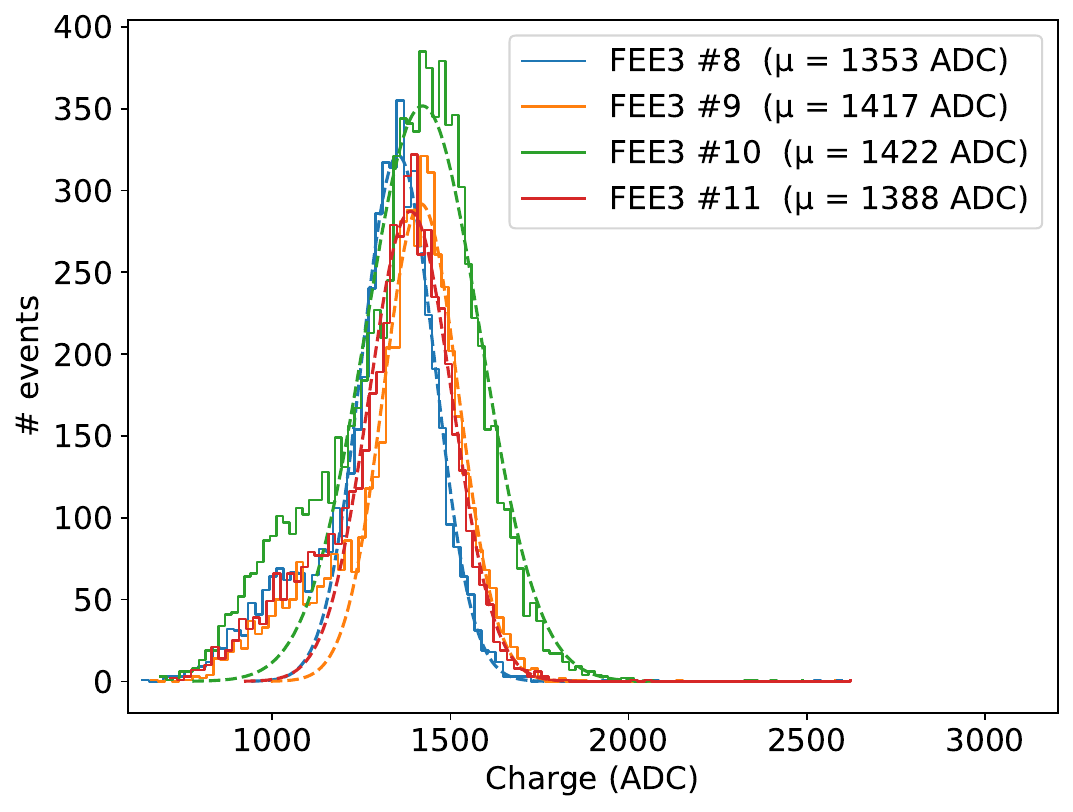}
\caption{\textbf{Left:} $^{55}$Fe spectrum measured with the 8 channels of an APD strip showing a very small spread as APDs are grouped by operating conditions. \textbf{Right:} Equalization of the APD gain through the bias high voltage, using the 5.9~keV line of $^{55}$Fe measured by the averaged channel on each board.}
\label{fig:apd_equalization}
\end{figure}

Before integrating the GAGG scintillators, the performance of each APD was verified individually using it as a direct X-ray detector, and its high-voltage bias was equalized across channels using the 5.9~keV line of $^{55}$Fe as a reference (see Figure~\ref{fig:apd_equalization}). Some noise issues were identified on the front-end electronics boards during this process, restricting the following absorber and coincidence tests to a subset of channels; the front-end board design will need to be optimized in the next iteration to make full use of all available channels.

\begin{figure}[htb]
\centering
\includegraphics[height=.38\textwidth]{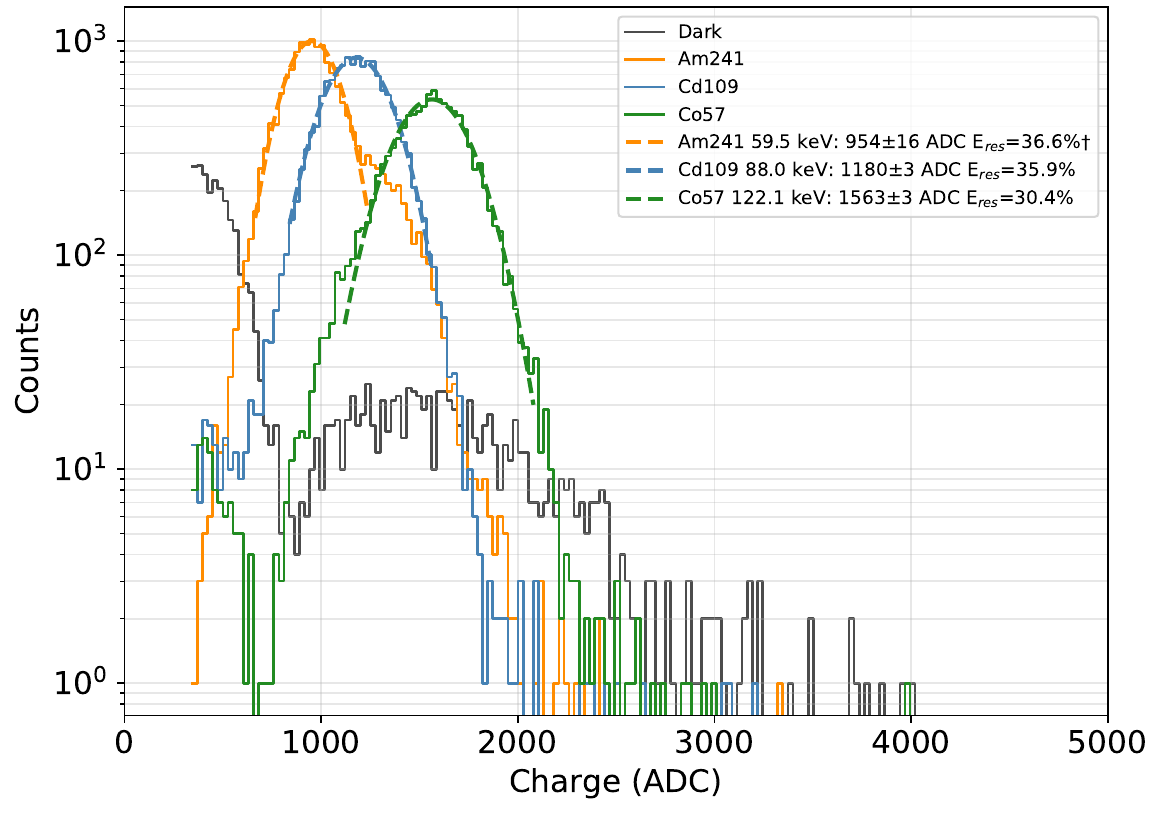}
\caption{Dark, $^{109}$Cd, $^{241}$Am, and $^{57}$Co spectra measured with an absorber channel. The photopeaks are fitted with Gaussians to extract the corresponding energy calibration and resolution shown in Figure \ref{fig:gagg_energy_calibration}.}
\label{fig:gagg_spectra}
\end{figure}

Once the APD gain was equalized and the GAGG scintillators integrated into the prototype, the absorber channels were energy-calibrated using the characteristic lines of $^{109}$Cd, $^{241}$Am, and $^{57}$Co sources. As shown in Figure~\ref{fig:gagg_spectra} for absorber channel 23, the 59.5, 88, and 122~keV lines from these sources could be clearly resolved and fitted with Gaussian profiles, while the 23~keV line of $^{109}$Cd remained buried below the noise level. This is the same light-collection issue already identified and discussed in detail for the plastic scatterers in Section~\ref{sec:scatterer_subsec}: the ESR reflective foils were glued directly onto the GAGG bars during assembly by Hilger Crystals, removing the air gap needed for total internal reflection at the scintillator surface and letting a large fraction of the scintillation light escape rather than reach the APD.

\begin{figure}[htb]
\centering
\includegraphics[height=.33\textwidth]{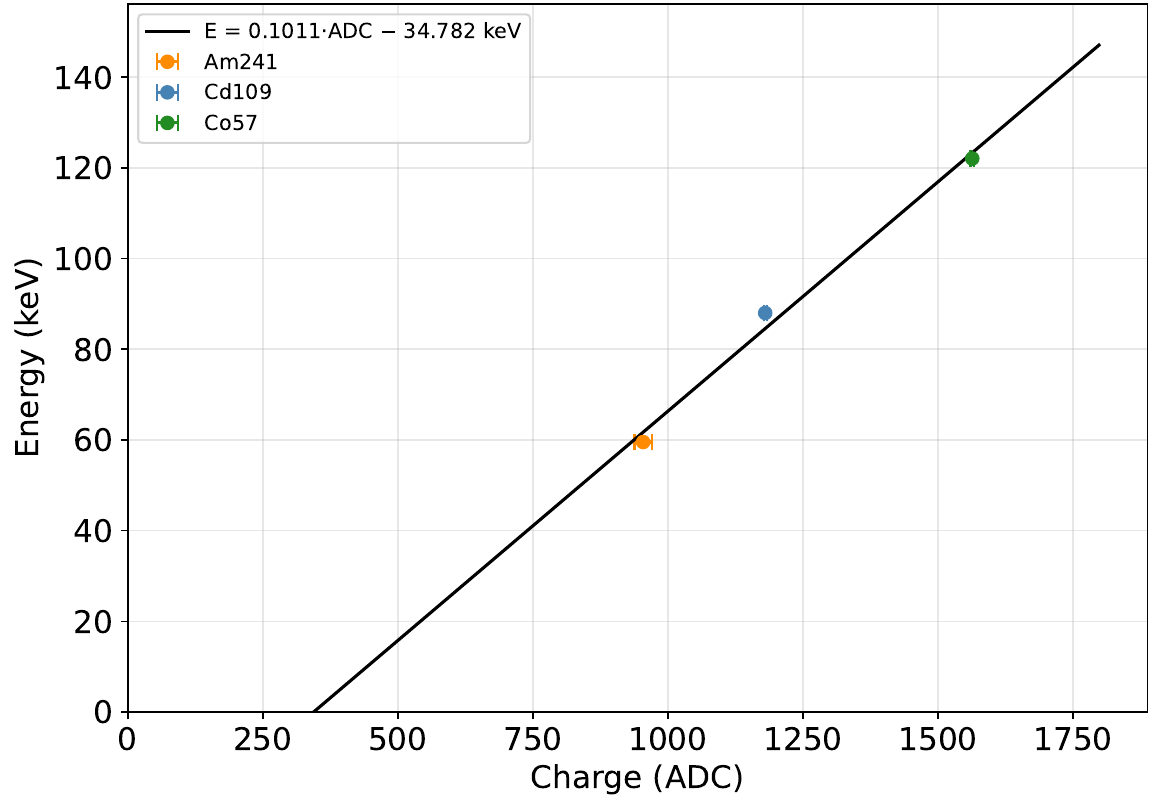}\hspace*{0.2cm}\includegraphics[height=.33\textwidth]{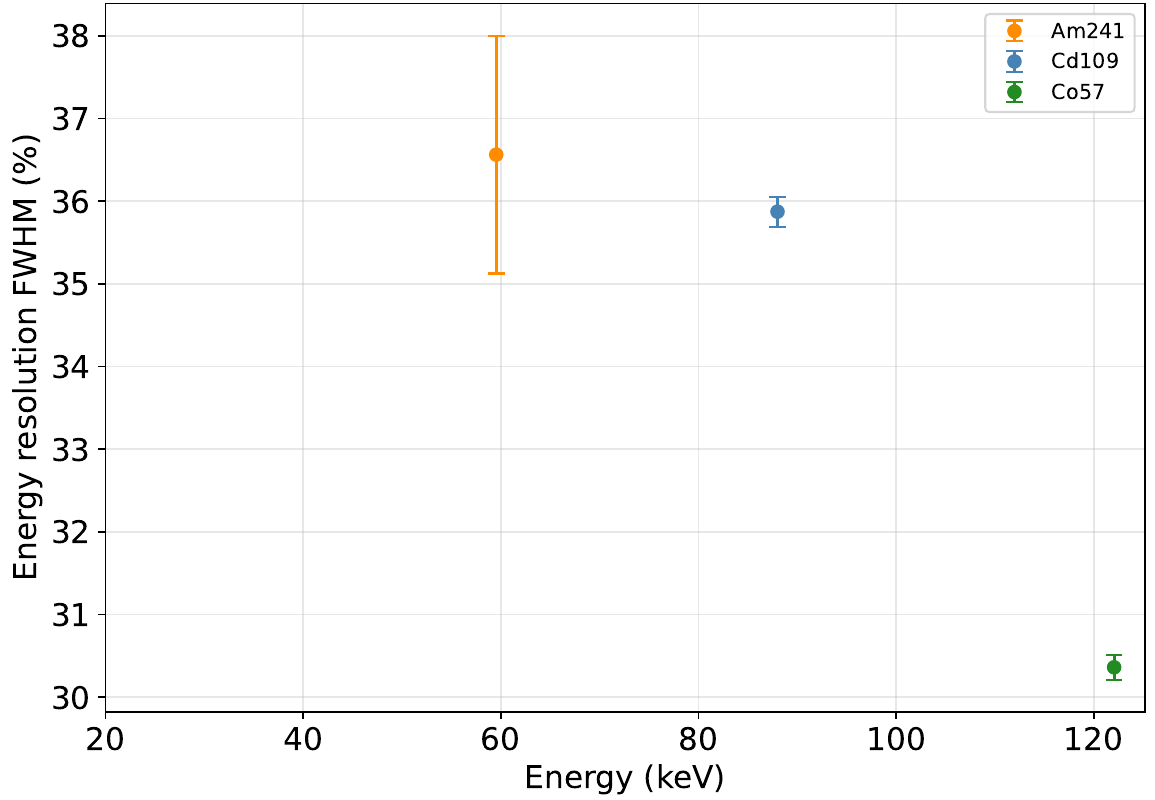}
\caption{\textbf{Left:} Energy of characteristic lines from $^{109}$Cd, $^{241}$Am, and $^{57}$Co sources as a function of the digitized charge. \textbf{Right:} Energy resolution of the absorber channel versus energy.}
\label{fig:gagg_energy_calibration}
\end{figure}

The measured charge of the resolved photopeaks as a function of the source energy (as shown in Figure~\ref{fig:gagg_energy_calibration}) shows the expected linear behavior, and the corresponding FWHM energy resolution (see Figure~\ref{fig:gagg_energy_calibration}) is well below the 50\% requirement at 59~keV. However, because of the reduced light yield discussed above, the 25~keV energy threshold required for the CUSP mission could not be verified with this prototype, and improving the light collection efficiency of the GAGG assemblies remains a priority ahead of the next prototype iteration.

\subsection{Coincidence measurements and spurious modulation}

\begin{figure}[htb]
\centering
\includegraphics[width=.98\linewidth]{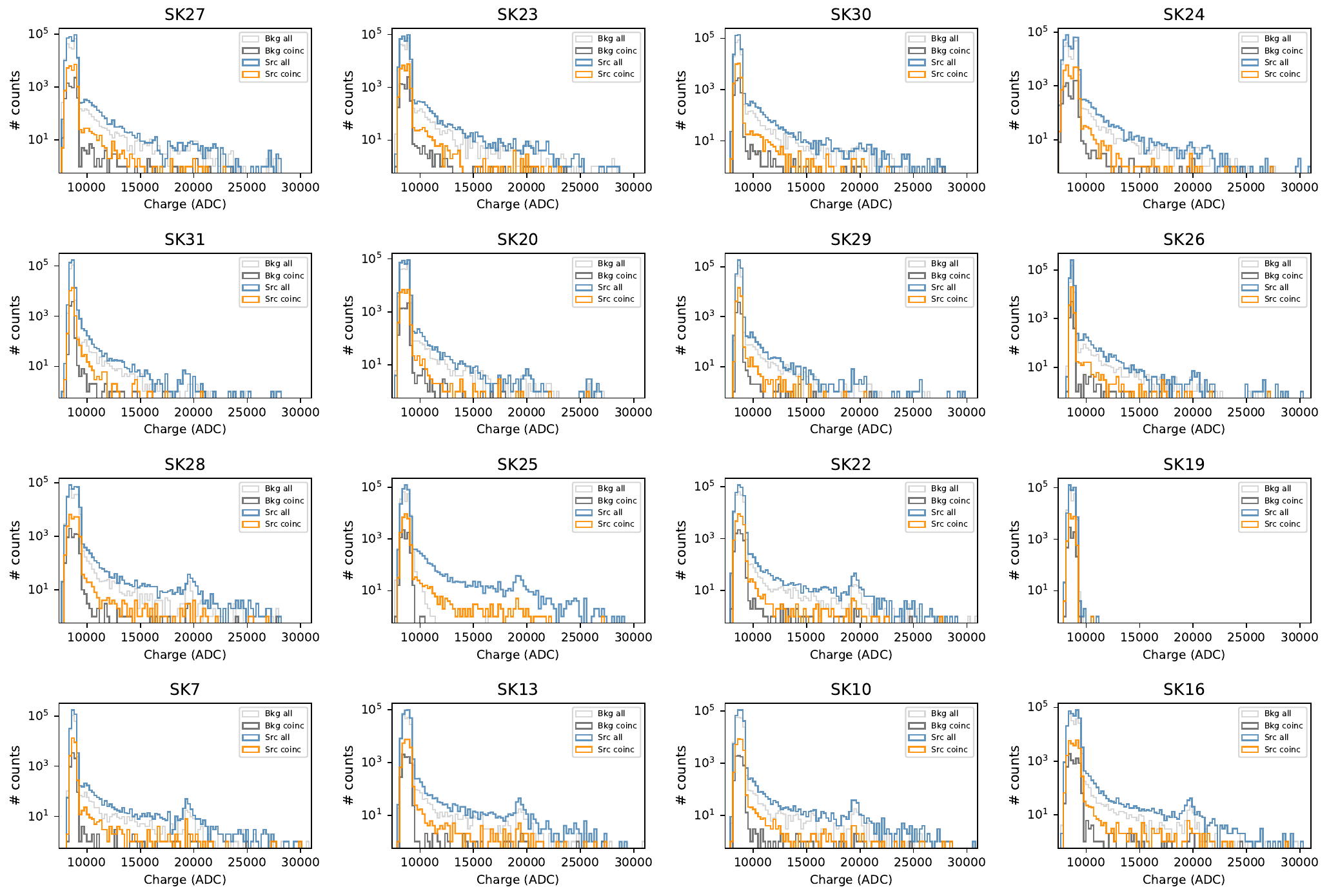}
\caption{Spectra measured by the 16 used absorber channels for background and unpolarized $^{57}$Co acquisitions both for coincidence-only and for all events.}
\label{fig:coincidence}
\end{figure}

With the gain-equalized and energy-calibrated prototype, coincidence measurements between the scatterer and absorber channels were performed, first on background and then illuminating a scatterer channel with an unpolarized $^{57}$Co source. The spectra measured for both acquisitions by all absorber channels with and without the coincidence mode are shown in Figure~\ref{fig:coincidence}. Reconstructing the azimuthal scattering angle of the coincident events allowed a first modulation curve to be built for an intrinsically unpolarized source, which is expected to translate into a uniform azimuthal distribution in the absence of instrumental effects.

\begin{figure}[htb]
\centering
\newsavebox{\rightcol}
\begin{lrbox}{\rightcol}
\begin{minipage}{0.48\linewidth}
\centering
\hspace*{-0.8cm}\includegraphics[width=1.08\linewidth]{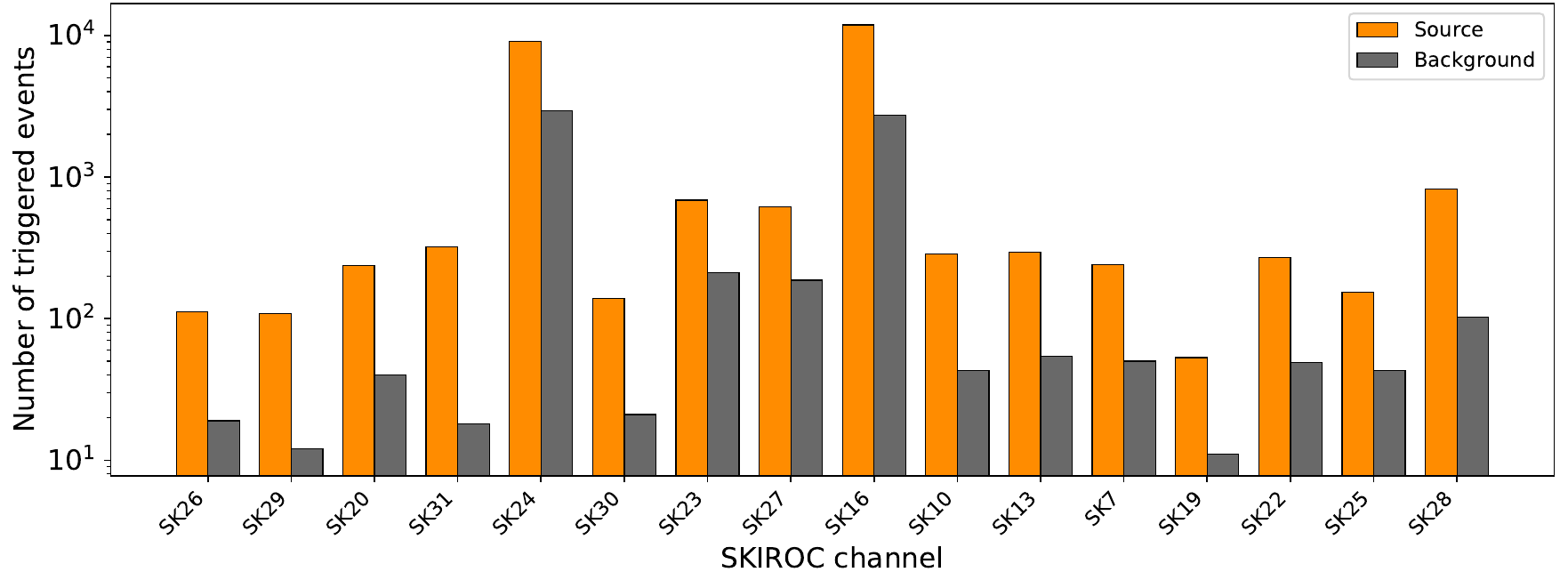}\\[0.3cm]
\hspace*{-0.8cm}\includegraphics[width=1.08\linewidth]{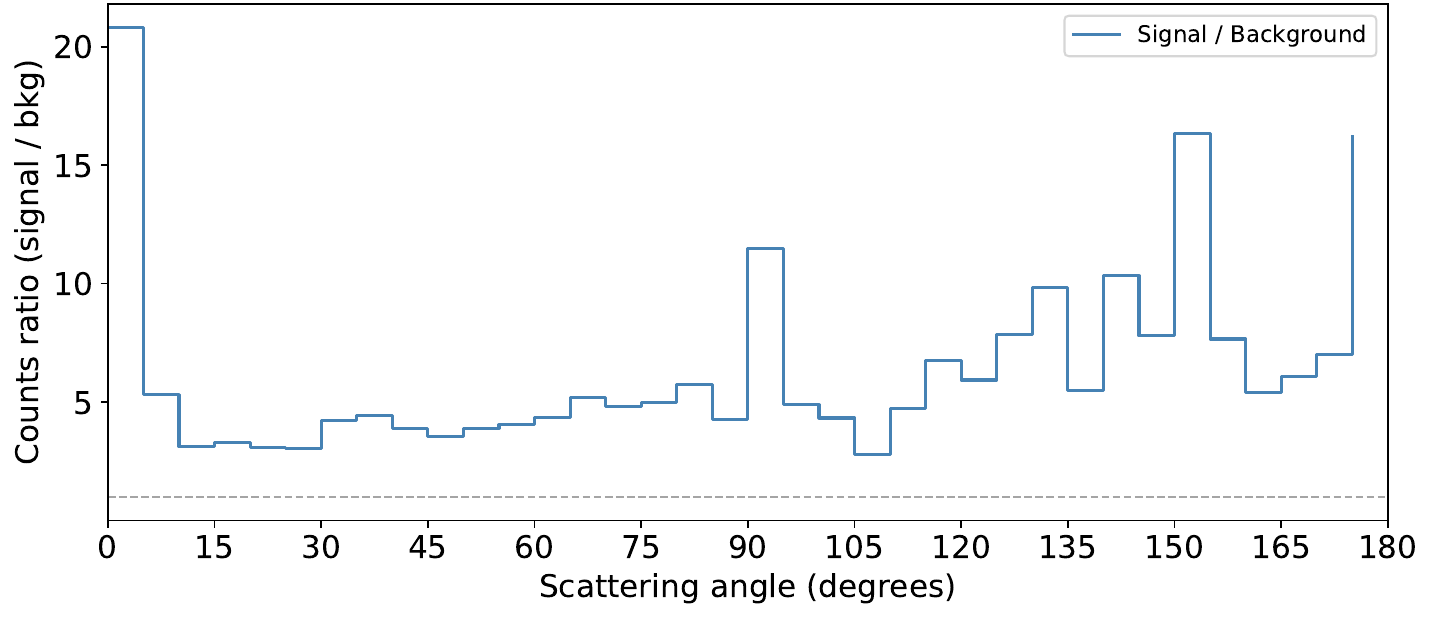}
\end{minipage}
\end{lrbox}

\newlength{\rightcolheight}
\settoheight{\rightcolheight}{\usebox{\rightcol}}
\newlength{\rightcoldepth}
\settodepth{\rightcoldepth}{\usebox{\rightcol}}
\addtolength{\rightcolheight}{\rightcoldepth}

\hspace*{-0.2cm}\begin{minipage}[c]{0.5\linewidth}
\centering
\includegraphics[height=\rightcolheight]{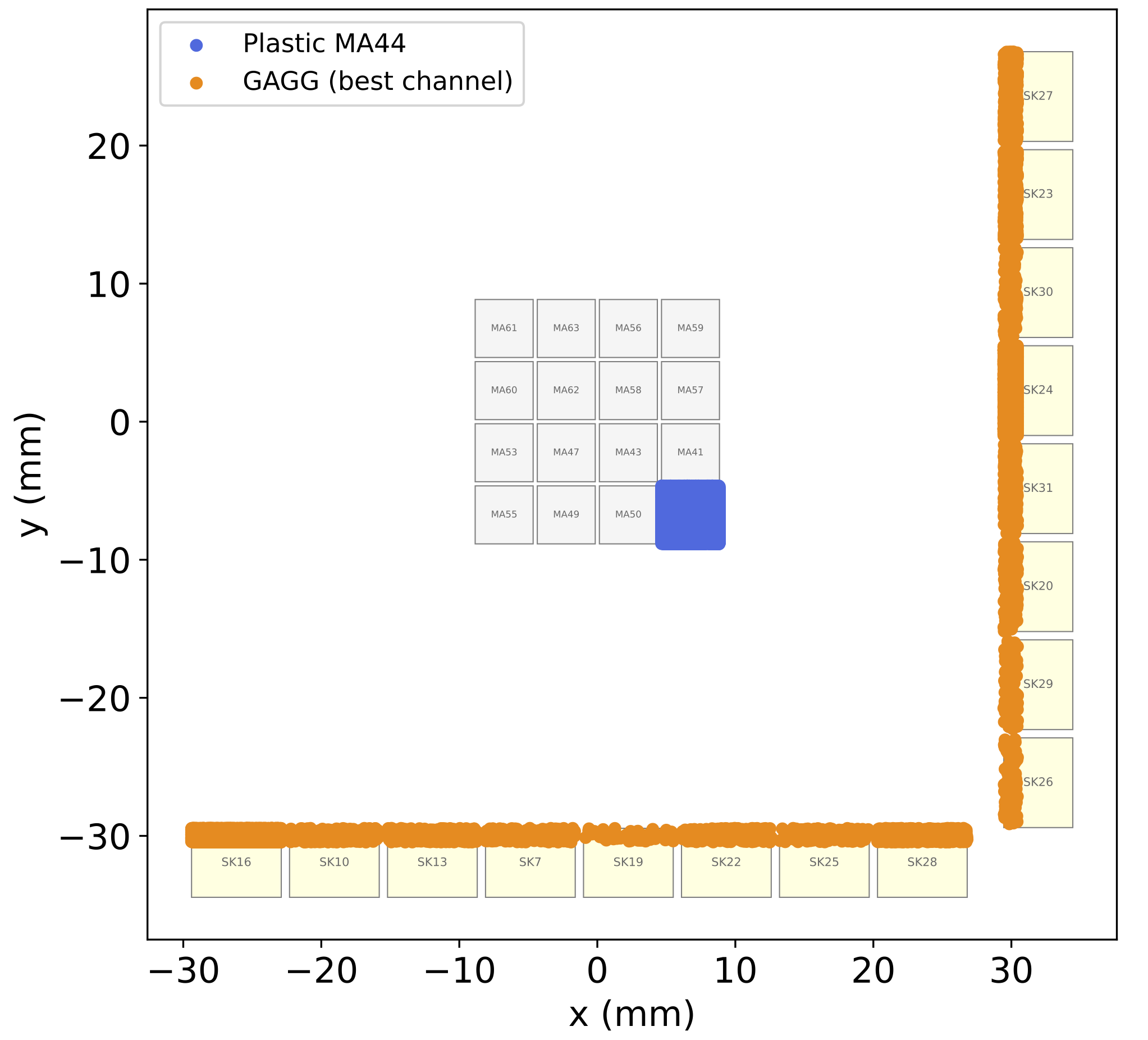}
\end{minipage}
\hfill
\usebox{\rightcol}
\caption{\textbf{Left:} Randomized hit map of the scattering and absorption positions for all coincidence events. \textbf{Top right:} Number of coincident events per absorber channel for the background and unpolarized source acquisitions. \textbf{Bottom right:} Azimuthal scattering angle distribution (modulation curve) measured with the unpolarized $^{57}$Co source, normalized to that obtained from the background measurement, showing the spurious modulation induced by noisy channels and non-uniform detector response.}
\label{fig:spurious}
\end{figure}

Instead, peaks are observed in the modulation curve obtained from the unpolarized $^{57}$Co measurement (Figure~\ref{fig:spurious}), which can be caused by noisy channels and a non-uniform response across the segmented detector. This systematic effect in the spurious modulation is significantly reduced when the unpolarized-source measurement is normalized by the corresponding background measurement, demonstrating that the residual non-uniformity can be effectively corrected for. More detailed measurements, including with polarized sources, will be performed on the larger-scale prototype to further characterize and minimize this systematic effect ahead of Phase C.

\newpage
\section{CONCLUSIONS AND OUTLOOK}
\label{sec:conclusions}

CUSP is a Compton polarimetry CubeSat mission aiming to significantly advance our understanding of solar flare physics and space weather forecasting through the first high-sensitivity, time-resolved measurements of the linear polarization of the non-thermal hard X-ray emission of solar flares. The mission, financed by ASI within the Alcor Program for CubeSat development, foresees a launch in late 2029/early 2030, pending approval of the successive project phases.

The Phase B of the project, started in December 2024, was concluded in July 2026. During this phase a first polarimeter prototype, based on a single central scattering block coupled to flight-representative front-end electronics, was designed, built, and characterized in the laboratory. This first characterization campaign confirmed the basic operating principle of the instrument, while also showing that the scintillator wrapping procedure and the front-end electronics board design will both need to be optimized to reach the energy resolution and channel yield required for the mission. The first modulation curve obtained with an unpolarized source also showed a clear spurious modulation pattern, which could nonetheless be significantly suppressed by normalizing to a background acquisition. Taken together, these results demonstrate that, once the identified light collection and electronics issues are addressed, the sensitivity required for solar flare polarimetry in the 25-100~keV band is within reach of the current design.

Addressing these findings is a central goal of the upcoming months, as we are moving towards phase C. A 30-month combined Phase C/D/E1 is being proposed to follow Phase B. The scintillator wrapping and optical coupling techniques will be revised to recover the light collection efficiency needed to reach the 25~keV threshold, guided by the optical characterization and simulation approach discussed in Section~\ref{sec:polarimeter} and in \cite{deangelis2025jinst}, while the various electronics board designs will be reviewed to eliminate the noise and manufacturing issues identified in this campaign. Further characterization of the polarimeter response, including with polarized sources and on the larger-scale 16-scatterer/32-absorber prototype, will be carried out in-house at INAF-IAPS using a new dedicated hard X-ray facility under construction based on the heritage of the IXPE Instrument Calibration Equipment\cite{muleri2022}.

\acknowledgments   
 
This work is funded by the Italian Space Agency (ASI) within the Alcor Program, as part of the development of the CUbesat Solar Polarimeter (CUSP) mission under ASI-INAF contract n. 2023-2-R.0.

\newpage
\bibliography{report} 
\bibliographystyle{spiebib} 

\end{document}